# Efficient Terahertz Generation from CoPt-based Terahertz Emitters via Orbital-to-Charge Conversion

Yongshan Liu[1,2,3]†, Yong Xu[1,2]†, Albert Fert[4]†*, Henri-Yves Jaffrès[4]†, Sylvain Eimer[1], Tianxiao Nie[2], Xiaoqiang Zhang[3], Weisheng Zhao[1,2,3]*

[1]National Key Lab of Spintronics, Institute of International Innovation, Beihang University, Yuhang District, Hangzhou, 311115, China

[2]Fert Beijing Institute, School of Integrated Circuit Science and Engineering, Beihang University, Beijing 100191, China

[3]Hefei Innovation Research Institute, Beihang University, Hefei 230013, China

[4] Laboratoire Albert Fert, CNRS, Thales, Université Paris-Saclay, 91767, Palaiseau, France

*Corresponding author: Albert Fert, Weisheng Zhao

Email: albert.fert@cnrs.fr, weisheng.zhao@buaa.edu.cn

†These authors contributed equally to this work

**Abstract**

Orbitronics devices operate by manipulating orbitally-polarized currents. Recent studies have shown that these orbital currents can be excited by femtosecond laser pulses in ferromagnet as Ni and converted into ultrafast charge current via orbital-to-charge conversion. However, the terahertz emission from orbitronic terahertz emitter based on Ni is still much weaker than the typical spintronic terahertz emitter. Here, we report more efficient light-induced generation of orbital current from CoPt alloy and the orbitronic terahertz emission by CoPt/Cu/MgO shows terahertz radiation comparable to that of efficient spintronic terahertz emitters. By varying the concentration of CoPt alloy, the thickness of Cu, and the capping layer, we confirm that THz emission primarily originates from the orbital accumulation generated within CoPt, propagating through Cu and followed by the subsequent orbital-to-charge conversion from the inverse orbital Rashba-Edelstein effect at the Cu/MgO interface. This study provides strong evidence for the very efficient orbital current generation in CoPt alloy, paving the way to efficient orbital terahertz emitters.

**Introduction**

In recent years, spintronic terahertz (THz) emitters based on ferromagnetic (FM)/non-magnetic (NM) heterojunctions have received broad attention due to their advantages such as large bandwidth, high compatibility, and low cost. Using femtosecond laser pulse to excite spin current in ferromagnets, the operating frequency of spintronic THz emitters extends to 30 THz[1,2]. The spintronic THz emitters utilize the spin current generated by ultrafast demagnetization and spin-charge current interconversion to generate efficient THz radiation[3,4]. Femtosecond pulses trigger the thermally driven ultrafast demagnetization of the FM layer, producing spin-polarized hot electrons[5,6]. These spin-polarized hot electrons propagate to adjacent metal layers and they are converted into transverse charge currents through mechanisms of spin-charge conversion, such as inverse spin Hall effect[7] or inverse Rashba-Edelstein effect[8,9]. Efficient spin-charge conversion is the key for developing advanced spintronic THz emitters with excellent and enhanced performances.

Recent work has found that when femtosecond lasers act on a ferromagnetic material as Ni, they generate orbital currents which are converted into in-plane charge currents and radiate THz waves[10,11]. The orbital-to-charge conversion can be obtained through the inverse orbital Hall effect (IOHE)[12] and/or inverse orbital Rashba-Edelstein effect (IOREE)[13]. These advances indicate that the use of orbital current can further enhance the emission efficiency of spintronic THz emitters. The IOHE and IOREE are the inverse effects of orbital Hall effect[14-19] and orbital Rashba-Edelstein effect[20-23], which have been found to exist in three-dimensional transition metals[24-28], at metal/oxide interfaces[29] or topological interfaces[30], and two-dimensional materials[31,32], etc.

At present, most of the orbital currents generated by femtosecond lasers have been obtained in the Ni-based system, but the Curie temperature and saturation magnetization of Ni are relatively low, making it unsuitable for developing spintronic THz emitters. Another approach for orbital current generation from a spin current is to insert a layer with high spin-orbit conversion efficiency characterized by a high spin-orbit correlation factor $\eta_{S-L}$[14, 16]. But this method introduces an additional conversion process, which limits the overall efficiency. Therefore, a suitable orbital material able to produce a direct large light-induced orbital currents still needs to be found. In this work, we demonstrate that a single-layer of CoPt alloy is an efficient source of orbital current. To characterize its orbital production, we designed a CoPt/Cu/MgO[29] structure and observed strong THz emission through femtosecond laser excitation experiments. By changing the thickness of Cu and the capping layer, we demonstrate that the THz emission of the system mainly originates from the light-induced orbital currents in CoPt and their orbital-to-charge conversion at the Cu/MgO interface, as previously demonstrated for this interface[11] and for other Cu/Oxide interfaces[24, 29] (Fig. 1(a)). Our work demonstrates the enormous potential of the orbital current of CoPt alloy in THz emission and paves the way for the development of efficient orbital THz emitters.

## Results and Discussion

**THz emission by light-induced orbital currents in CoPt alloys**

We used a magnetron sputtering system to deposit CoPt (3 nm)/Cu (2 nm)/MgO (2 nm) on SiO$_2$ substrate, with several relative concentrations of Co and Pt in CoPt. We quantified these relative concentrations by using energy dispersive X-ray spectroscopy (EDS) (Fig. 1(b)). In our nominal Co$_{50}$Pt$_{50}$ alloy, the compositions of Co element (blue) and Pt element (red) in EDS images are 49.30% and 50.70%, respectively. Both elements show a uniform distribution. Fig. 1(c) shows the X-ray diffraction (XRD) results of the sample, indicating that the CoPt layer deposited on an amorphous SiO$_2$ substrate shows (111)-orientation. We use femtosecond lasers to pump samples and obtain THz spectra using a THz time-domain spectroscopy system with experimental details given in the Methods section. Fig. 1(d) shows the THz waveform of CoPt/Cu/MgO, indicating that

CoPt/Cu/MgO generate efficient THz emission, with a peak electric field intensity ($E_{peak}$) close to that of the typical spintronic THz emitter Co (3 nm)/Pt (2 nm), also prepared for direct comparison.

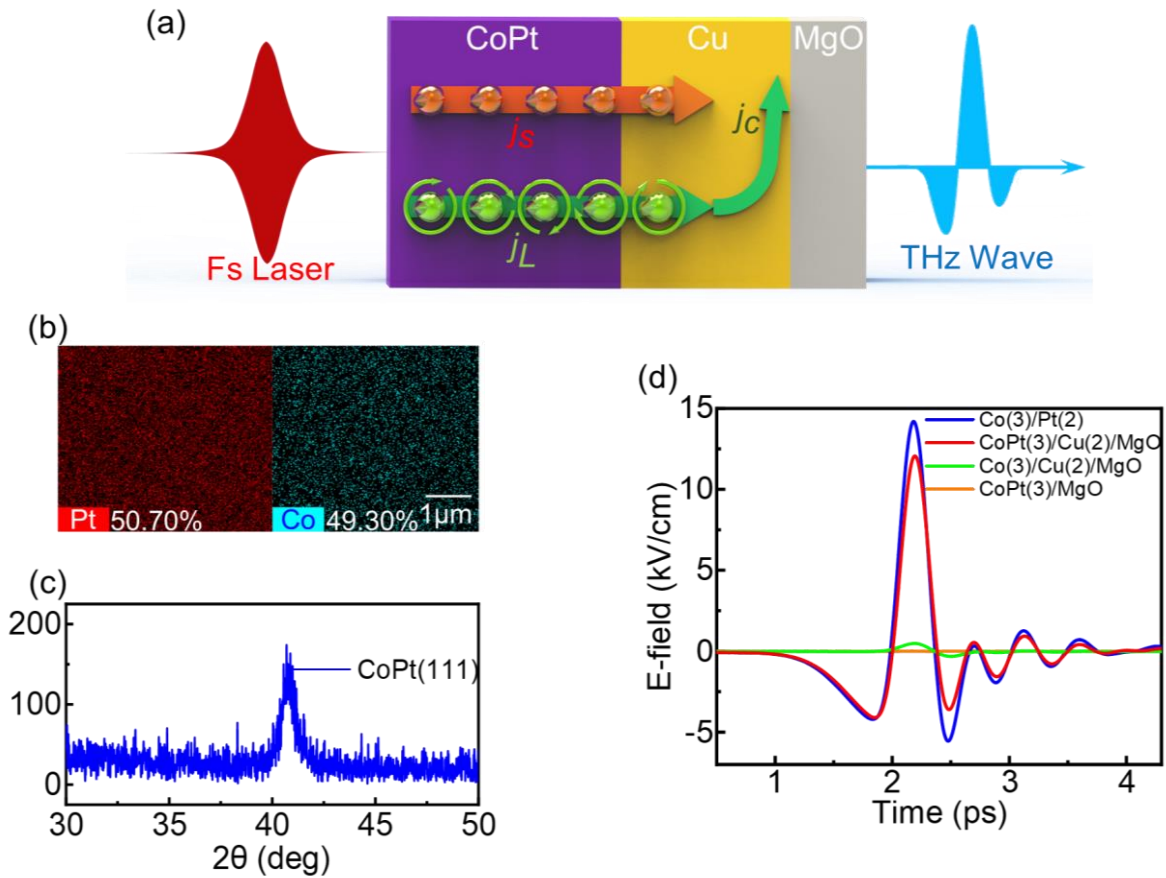

Fig. 1(a) Schematic illustration of THz emission generated from CoPt/Cu/MgO, the laser excites CoPt to generate $j_s$ and $j_L$ in the situation where only $j_L$ is converted into a charge current $j_c$ after injection into the Cu layer, and radiates THz wave. (b) The EDS image of CoPt, with Co element in blue and Pt element in red, showing uniform distribution, according to EDS, Co is 49.30% and Pt is 50.70%. (c)The XRD shows that CoPt is (111)-oriented. (d) The THz spectra of Co/Pt, CoPt/Cu/MgO, Co/Cu/MgO, and CoPt/MgO measured under the same conditions, showing that the $E_{peak}$ of CoPt/Cu/MgO is close to the $E_{peak}$ of Co/Pt, which is much stronger than that of Co/Cu/MgO. In addition, no THz is generated in the CoPt/MgO, the numbers in parentheses represent the thickness.

In order to identify the THz emission by our CoPt/Cu/MgO heterojunctions in Fig. 1(d), we first measured the THz radiation of CoPt (3 nm) single layer and could not detect any sizeable THz signal. THz emission from single-layer ferromagnets has been already observed and attributed to the anomalous Hall effect (AHE) of the ferromagnetic layer[33-35]. The absence of signal with a CoPt single layer proves that the THz emission of CoPt/Cu/MgO is e.g. not related to the AHE of CoPt. Another possible mechanism is the light-induced generation of out-of-equilibrium spin accumulation and subsequent spin current $j_s$, as illustrated by the red arrow in Fig. 1(a). To check whether the THz emission of CoPt/Cu/MgO can have a contribution from spin currents, we prepared a sample in which CoPt was replaced by Co for in which the emission of light-induced spin current is well-known[36]. As shown in Fig. 1(d), we found that the THz radiation of Co/Cu/MgO was only 4% of the radiation observed in CoPt/Cu/MgO. This result indicates that, in the FM/Cu/MgO system and as expected from the negligible spin Hall angle of Cu, the charge current generated by spin and spin-charge conversion can only be small and cannot explain the strong THz radiation observed in CoPt/Cu/MgO. After excluding contributions from AHE in CoPt layer THz effect and from spin currents, we conclude that the THz emission of CoPt/Cu/MgO originates from orbital effects. When a femtosecond laser acts on a heterojunction, the light-induced orbital accumulation $\mu_L$ generated inside CoPt is converted into an orbital current $j_L$ through Cu and finally into a charge current $j_c$ via the orbital-to-charge conversion effect, generating THz emission, as represented by the green arrow in Fig. 1(a). In the following, we will refer indifferently to the orbital accumulation $\mu_L$ and orbital current $j_L$ both propagating inside Cu, one quantity being simply linearly proportional to the other from their same exponential spatial dependence $\mu_L = e^{-\frac{d_{Cu}}{\lambda_{\phi l}}}$ if one neglects the carrier reflection at Cu/MgO interface ($\lambda_{\phi l}$ is the orbital decoherence length).

In order to further investigate the source of orbital accumulation and orbital current, we prepared several different alloy compositions of CoPt/Cu/MgO thin films and studied the effect of Pt doping on THz emission. Fig. 2(a) shows that, as the Pt component increases, the $E_{peak}$ rapidly increases (Fig. 2(b)). In the absence of Pt doping, we found weak THz emission in Co/Cu/MgO, while single-layer Co (3 nm) showed no significant THz emission. These results suggest that Pt doping plays a decisive role in the generation of orbital density and orbital currents. This may be related to the strong spin-orbit coupling of Pt element[37-39], either by direct generation of orbital polarization during the de-excitation process or related to the $\eta_{S-L}$ conversion factor defined previously and increasing significantly with the increase of Pt doping.

In addition, the signal with CoPt shows a significant shift in time, presently exceeding 40 fs in $Co_{50}Pt_{50}$, as shown in Fig. 2(c). It indicates that there is a delay time in the generation of charge current, similar to the delay in recent reports on ballistic transport in orbital THz emitters[10, 11]. Consistently, according to a recent literature, it is suggested that such delay time is more characteristic of the orbital transport (through Cu) than spin transport[40].

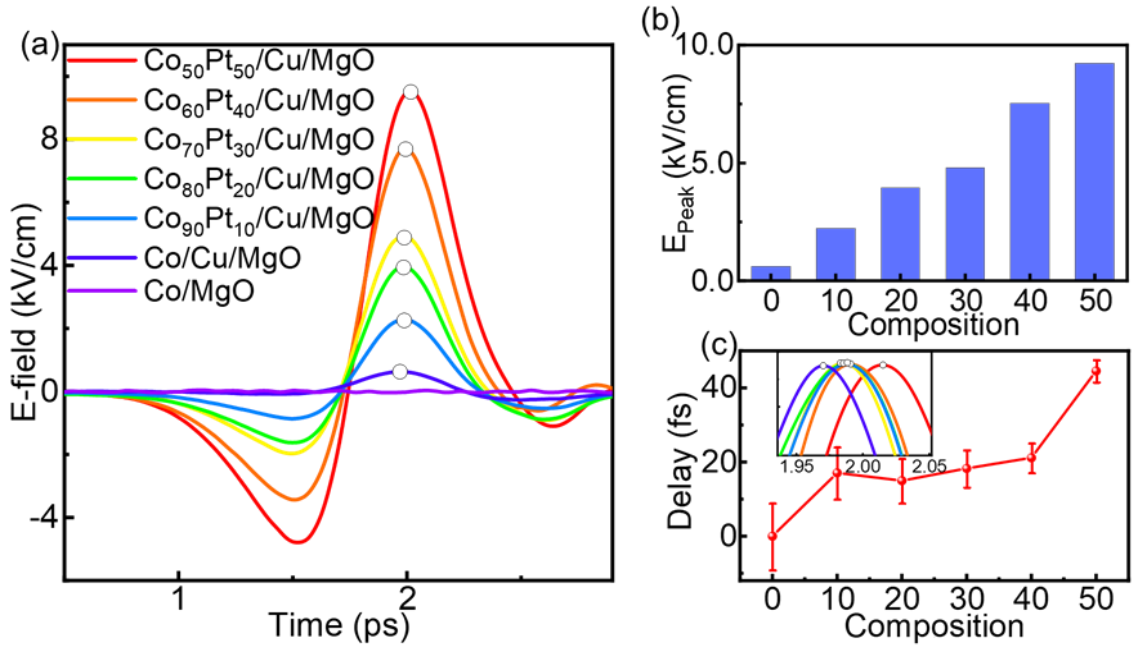

Fig. 2(a) THz spectra of CoPt with different components. With the increase of Pt doping, THz emission gradually increases, and the THz wave gradually shifts positively. The (b) $E_{peak}$ values and (c) delay time extracted from Fig. 2(a), where the delay time of $Co_{50}Pt_{50}$ exceed 40 fs. The inset shows the normalized waveform.

**Orbital-to-charge conversion at Cu/MgO interface.**

The previous results indicate that the THz emission originates from the orbital current in CoPt and its conversion into charge current in Cu (by IOHE) or at the interfaces of Cu (by IOREE). Whereas a very small OHE of Cu is theoretically predicted[41], efficient OREE conversions have been already observed at interfaces of Cu and oxides, for example by Kim et[24, 29]. To check that the IOREE at the Cu/MgO interface is the predominant conversion mechanism, we prepared and tested CoPt/Cu/$Si_3N_4$ samples under the same conditions. The $E_{peak}$ values of CoPt/Cu/MgO and CoPt/Cu/$Si_3N_4$ are shown in Fig. 3 with, as expected, inverted signal for opposite field directions. When MgO is replaced by $Si_3N_4$, THz emission decreases by 80%. This result indicates that the large contribution of THz emission with CoPt/Cu/MgO comes from the efficient orbital-to-charge conversion by IOREE at the Cu/MgO interface. The recent studies on charge-to-orbital conversion at the Cu/oxide interface have ascribed the efficiency of the conversion to oxygen atom diffusion at the oxide interface[22, 29, 42]. Our results confirm these results and further demonstrate that, through the orbital-to-charge conversion at the Cu/MgO interface, the orbital current generated from CoPt is converted into a charge current, which constitutes the main contribution to THz emission.

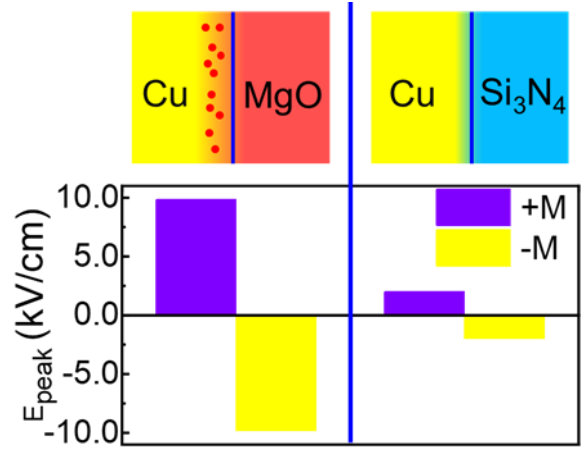

Fig. 3 $E_{peak}$ values from different interfaces: Cu/MgO, Cu/Si$_3$N$_4$, The $E_{peak}$ value of Cu/Si$_3$N$_4$ is only 20% of the of Cu/MgO. The purple and yellow correspond to the positive and negative fields respectively.

**Ballistic orbital transport, orbital carrier velocity and decoherence time in Cu.**

The interpretation of most experiments in orbitronics are based on a diffusive propagation of the orbital current characterized by an orbital diffusion length similar to the spin diffusion length in spintronics[27, 43, 44]. The situation is different in experiments of THz emission generated by light induced orbital current in the conditions of ballistic transport, that is *i)* with mean free path $\lambda_{Cu}$ (distance between successive scatterers) longer than the thickness $d_{Cu}$ of the spacer layer (Cu in our sample); or *ii)* in the case of a characteristic momentum relaxation time $\tau_p$ longer than the characteristic decoherence time $\tau_{\phi l}$. We can refer to Seifert et al for a precise description[10] of the ballistic regime and its transition to the diffusive regime. The ballistic regime is characterized by a certain time delay of the THz signal increasing with the spacer thickness $d_{Cu}$ as $\frac{\sqrt{3}d}{v_l}$, with $v_l$ is the group velocity of the orbital carriers. The description of the ballistic regime has been later improved by considering the effects of an intrinsic decoherence time of the orbital carriers (not related to scattering and without interplay with the ballistic character), $\tau_{\phi l}$ in reference[11], similar to the scattering independent lifetime

$\tau$ introduced by Go et al[26]. In our Supplementary Material (Section 1), we present a global description of the properties of the ballistic regime, that is, mainly, the existence of the delay characteristic of the ballistic regime, $\tau_D = \frac{\sqrt{3}d}{v_l}$ with an emission amplitude varying as $exp\left(-\frac{d}{\lambda_{\phi l}}\right)$ with $\lambda_{\phi l} = \frac{v_l}{\sqrt{3}}\tau_{\phi l}$.

In Fig. 4(a), we present the THz emission waveforms of CoPt/Cu/MgO samples with different Cu thicknesses. The THz emission, already large at 1 nm, reaches its maximum value when the Cu thickness is 2 nm (that we identify to the thickness for a complete coverage of Cu on MgO), and gradually decreases with increasing the Cu thickness ($d_{Cu}$). The THz emission at 4 nm is only 25% of that at 2 nm. The rapid decrease of signal can be due to several factors acting in parallel involving notably *i*) a decrease of the light absorption in the CoPt layer, *ii*) a decrease of the film impedance[1], and *iii*) a loss of orbital current. In order to quantitatively analyze the orbital transport, we used the transfer matrix method (TMM) proposed by Yang et al. to study the THz emission for a given heterojunction structures[45] considering the absorption process *i*). This method effectively simulates the propagation and emission of THz electromagnetic waves in heterostructure devices by introducing the free space distribution of current and has been successfully used to analyze the spin-charge conversion process of Fe/Au[45]. In order to analyze the THz emission efficiency, the light absorption in CoPt/Cu/MgO has to be first removed (see Supplementary Section 2) so as to highlight the intrinsic behavior of the orbital transport through Cu giving rise afterwards to the orbital-charge conversion at Cu/MgO interface. For that goal, we used a phenomenological exponential decay model to describe the attenuation of orbital current and accumulation inside Cu according to $\mu_L = e^{-\frac{d_{Cu}}{\lambda_{\phi l}}}$ (Fig. 4(b)), where $\lambda_{\phi l} = \frac{v_l}{\sqrt{3}}\tau_{\phi l}$ represents the typical ballistic decoherence length (see Supplementary Section 1B for more explanation). Likewise, we successfully managed to fit the intrinsic Cu thickness dependence by considering $\lambda_{\phi l} \approx 2.9$ nm giving thus a pretty close estimation of the orbital decoherence length (Fig. 4(c)).

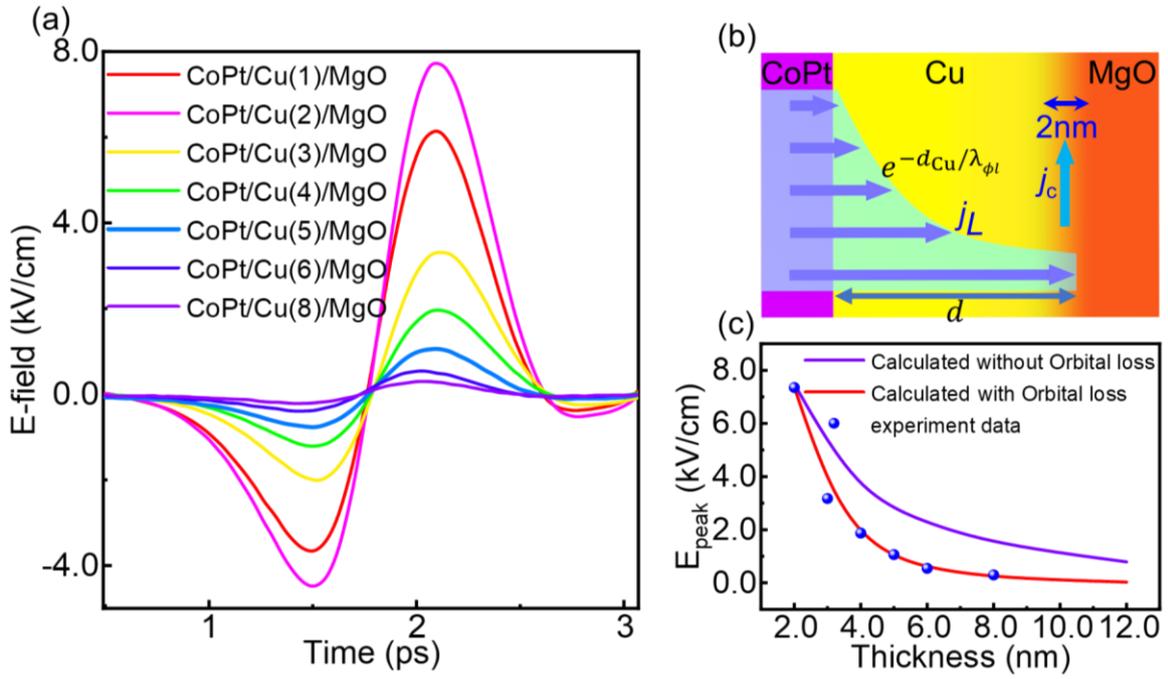

Fig. 4(a) Cu thickness dependence of THz waveform from CoPt/Cu/MgO emitter, the maximum is reached when Cu is 2 nm, and then decreases rapidly with the increase of thickness. (b) Schematic illustration of $j_L$ transport in Cu, which reaches the Cu/MgO interface after an exponential decay and converts to $j_c$ within the 2 nm region. (c) Using TMM to fit $E_{peak}$ as a function of thickness, the blue dots representing the measured data. The red curve (purple curve) representing the fitting results by the TMM with (without) taking into account exponential decay.

Based on the above analysis, the noticeable delay time of CoPt/Cu/MgO in Fig. 2(c) can be interpreted as the orbital transit time through Cu before orbital-to-charge conversion at the Cu/MgO interface. The delays in CoPt/Cu(x)/MgO were extracted and their variation upon the Cu thickness is represented by the blue line in Fig. 5(a). When the Cu thickness increases, the signal first shifts in time positively and the corresponding delay displays a characteristic linear variation in $d_{Cu}$, as expected for the ballistic regime. However, it reached a maximum value only when Cu was 3 nm, after which the signal starts to shift negatively by 90 fs at 6 nm from the maximum. The positive shift of the signal (positive delay) is characteristic of a non-local orbital transport and has to be assigned to a ballistic transport of $\mu_L$ (see Supplementary Section 1). The estimated orbital group velocity is $\frac{v_l}{\sqrt{3}} \sim 0.15$ nm/fs, according to the values in Fig. 5(a).

The fast decrease of the time delay beyond 3 nm Cu witnesses a well shorter transit time of the information transported. It may reveal a crossover between orbital and spin-charge conversion process for quite thick Cu and predominant spin current contribution. To validate this hypothesis, Co/Cu(x)/MgO series samples, free of any orbital contribution, were measured and the extracted delay, displayed by a red line in Fig. 5(a), displays almost no variation. According to Gueckstock et al., the THz of 3d FM (=Py, Co, Fe)/Cu bilayers is mainly generated from the spin-charge conversion occurring locally at the FM/Cu interface[36]. This mechanism at FM/Cu interface seems to be supported here by our experimental observations in Co/Cu owing to *i*) the delay time does not vary with the Cu thickness; *ii*) and the THz intensity of Co/Cu/MgO does not decrease significantly with the Cu thickness (see Supplementary Section 3). Based on these observations, we propose the following explanation sketched in Fig. 5(b). For Cu (> 3 nm), the charge current generated by orbital current, in comparison with the charge current generated by spin current becomes negligible due to the relative short decoherence length of $\mu_L$ in Cu. This makes apparent the charge current with about zero delay generated from the spin and caused by skew scattering locally at the CoPt/Cu interface, resulting in a negative shift of the waveform.

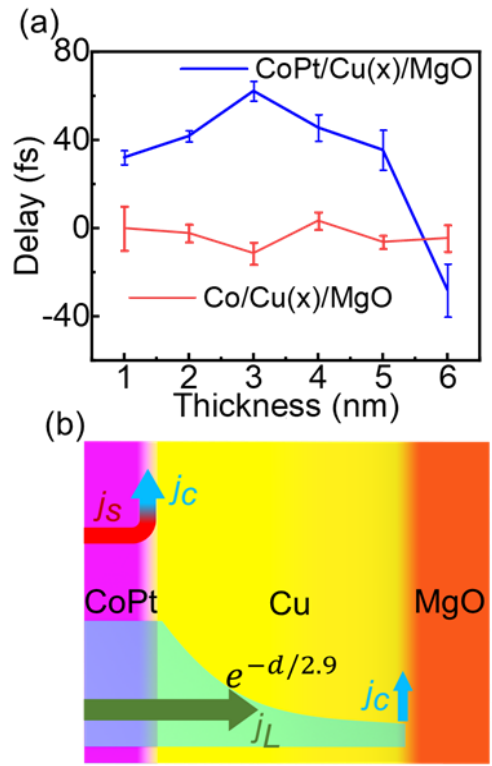

Fig. 5(a) The signal shifts of CoPt/Cu and Co/Cu vary with the thickness of Cu, as the thickness of Cu increases, CoPt/Cu first shifts positively and then shifts negatively, while Co/Cu does not show a significant shift. (b) Schematic illustration of $j_L$ and $j_s$ transport in Cu, $j_L$ reaches the Cu/MgO interface after an exponential decay and $j_s$ triggers spin-charge conversion at CoPt/Cu interface.

**Conclusion**

In summary, our experiments show the very efficient THz generation by light-induced orbital currents in CoPt/Cu/MgO, at the level of the best spintronic THz emitters. CoPt turns out as an efficient source of orbital current, efficiency associate with the introduction of the large SOC of Pt in Co. Ni was already known for the production of light-induced orbital current[10, 11] but the CoPt alloys appear to be much more efficient sources of orbital current. The conversion from orbital-to-charge current generating the THz emission is confirmed to occur at the Cu/MgO interface, induced by the IOREE as already observed in other Cu/Oxide interfaces. The propagation of ballistic orbital currents through Cu has been quantitatively analyzed, revealing an orbital decoherence length in Cu of only 2.9 nm, significantly smaller than the spin diffusion length in metals[44]. It is noteworthy that THz emission from CoPt/Cu/MgO displays a pronounced time delay, indicative of the clear ballistic transport nature of $\mu_L$ within Cu. Our findings on this delay offer time-resolved insights into the transport of orbital currents in metals.

We found that an emission from light-induced orbital current in CoPt can be almost as large than the emission by the best spintronic emitters based on spin-to-charge conversion by heavy metals. CoPt is promising from the application point of view, especially if one considers of the potential of further optimized devices with orbital and spin contributions are judiciously added, as we have observed in a preliminary result on a CoPt/Pt sample. This will broaden the scope of research for THz emitters.

**Methods**

**Sample growth.** CoPt alloy layer was deposited on an SiO$_2$ substrate using Co and Pt targets for co-sputtering at room temperature. The base pressure was maintained at $1\times10^{-9}$ mbar, and the Ar gas pressure of $5\times10^{-3}$ mbar was applied. Subsequently, Cu and MgO were deposited in sequence.

**THz measurement**. The THz emission spectrum was performed in a home-built THz time-domain spectrometer system. The femtosecond laser pulse is generated by an amplified Ti:sapphire laser with the repetition frequency of 1 kHz. The central wavelength of the laser pulse is 800 nm and the duration of the pulse is 35 fs. The THz pulse is detected by electro-optical sampling method using a 0.5 mm thick ZnTe crystal.

The THz electric field can be obtained from the measured electro-optic signal sampled on the ZnTe crystal. Based on this, the relationship between the THz electric field and the voltage measured on the balanced detector can be expressed by the following equation[46]:

$$E_{\text{THz}} = \frac{2c}{\omega n^3 r_{41} L} (\cos\alpha \sin 2\beta + 2\sin\alpha \cos 2\beta) \left(\frac{\Delta V}{V}\right) (V/cm) \quad (1)$$

Where $c = 3\times10^8$ m/s is the speed of light in vacuum, $\omega = 2\pi c/800$ nm is the angular frequency of the probing light, $n = 2.85$ is the refractive index of ZnTe (110), $r_{41} = 3.9\times10^{-12}$ is the electro-optic tensor, and $\alpha$ is the angle between the THz polarization and the (001) axis, and $\beta$ is the angle between the probing light polarization and the (001) axis, both of which are 90°. $\Delta V/V$ represents the differential change in the intensity of the probing light measured by the detector.